\documentclass[12pt]{article}
\usepackage{amsmath}
\usepackage{amssymb}
\begin{document}
\begin{flushright}
KFKI-1985-28\\
HU ISSN 0368 5330\\
{\it ISBN 963 372 361 2}\\
March 1985
\end{flushright}
\vskip 3truecm
\begin{center}
ORTHOGONAL JUMPS OF WAVEFUNCTION\\ IN WHITE-NOISE POTENTIALS
\vskip 1truecm
L. Di\'osi
\vskip .5truecm
Central Research Institute for Physics\\
H-1525 Budapest 114, P.O.B. 49, Hungary
\end{center}
\vskip 1truecm
ABSTRACT

We investigate the evolution of the quantum state for a free particle 
placed into a random external potential of white-noise type. The master
equation for the density matrix is derived by means of path integral method. We
propose an equivalent stochastic process where the wavefunction satisfies a
nonlinear  Schr\"odinger equation except for random moments at which it shows
orthogonal jumps. The relation of our work to the usual theory of quantum
noise and damping is briefly discussed.
\newpage

Since the early works [1,2,3] on the theory of quantum noise
and damping, a great interest has been shown in quantummechanical
systems affected by random forces and also, new aspects have
appeared [4,5,10]. Here we shall investigate the effect of Gaussian
white-noise potentials.

For brevity we restrict ourselves to the case of a single
point-line particle moving in one dimension.

Let us assume that the potential $V(x,t)$ acting on the particle
is a stochastic variable ($x,t$ stand for the coordinate and
time). We define the probability distribution of $V$ by the following
generator functional $G[h]$:
$$
G[h]\equiv\langle\exp (i\int V(r,t)h(r,t)drdt)\rangle=
$$
\vskip -1 truecm
$$
\eqno{(1)}
$$
\vskip -.5 truecm
$$
~~~~~~~~~~~~~~~~=\exp(-\frac{1}{2}\int h(r,t)h(r^\prime,t)f(r-r^\prime)drdr^\prime)
$$
where $h$ is an arbitrary function. The symbol $\langle\rangle$ stands for expectation
values evaluated by means of the probability distribution
of $V$. Functional differentiation of $G[h]$ gives rise to the two
tipical relations of moments of white-noise type:
$$
\langle V(r,t)\rangle=0,
\eqno{(2)}
$$
$$
\langle V(r,t)V(r^\prime,t^\prime)\rangle=\delta(t-t^\prime) f(r-r^\prime).
\eqno{(3)}
$$

For further use, we introduce tho following notation:
$$
g(r)=f(0)-f(r).
\eqno{(4)}
$$

Now we investigate the effect of the white-noise potential
(1) on the quantummechanical motion of a given point-like particle
of mass $m$.

If we single out a given potential $V$ then the wavefunction
$\psi_t(x)$ of the particle will  satisfy the Schr\"odinger equation of
motion:
$$
\frac{\partial}{\partial t}\psi_t(x)=\frac{i\hbar}{2m}\frac{\partial^2}{\partial x^2}\psi_t(x)-\frac{i}{\hbar}V(x,t)\psi_t(x)
\eqno{(5)}
$$
where $x$ is the spatial coordinate and $t$ refers to the actual value
of the time. Taking the initial wavefunction $\psi_0(x)$ at $t=0$, one
can express the solution for $\psi_t(x)$ by means of the Feynman's
path integral formula [7]:
$$
\psi_t(x)=\int\exp\{\frac{i}{\hbar}\int\displaylimits_0^t[\frac{m}{2}{\dot x}^2_\tau-V(x_\tau,\tau)]d\tau\}\psi_0(x_0)Dx_\tau.
\eqno{(6)}
$$
\vskip -1.1 truecm
$$
~~~~~~~~~~~~~~~~~~~~~~~~~~~~~~~~~~~~~~~~~~~~~~~~~~~~~~~~~~~~~~~~t\!>\!\tau\!\geq\!0
$$

In our case,  $V(x,t)$ is a stochastic variable, thus $\psi_t(x)$ will
evolve in time according to a give stochastic process. We shall
not derive the rules of this stochastic process. Instead, we shall
construct another stochastic process for $\psi_t(x)$ which leads to the
correct statistical predictions and which has genuine features
from the viewpoint of measurement theory.

In the generic case, the quantum state of the particle is
uniquely characterized by the density matrix [6]
$$
\rho_t(x,y)\equiv\langle\psi_t(x)\psi^{\!\!\!\ast}_t(y)\rangle.
\eqno{(7)}
$$

Indeed, it can be shown [8] that $\rho_t$ yields all the usual
statistical predictions of the quantummechanics. Namely,
$$
O_t=\int\widehat O(y,x)\rho_t(x,y)dxdy
\eqno{(8)}
$$
where $\widehat O$ is the hermititian operator of an arbitrarily given dynamical
quantity and $O_t$ stands for its predicted value at time $t$.

First, we shall prove that the density matrix (7) satisfies
a parabolic differential equation of motion. Using eq. (6) along
with eq. (7), one gets:
$$
\rho_t(x,y)=~~~~~~~~~~~~~~~~~~~~~~~~~~~~~~~~~~~~~~~~~~~~~~~~~~~~~~~~~~~~~
$$
$$
=\langle\int\exp\{\frac{i}{\hbar}\int\displaylimits_0^t[\frac{m}{2}({\dot x}^2_\tau-{\dot y}^2_\tau)-
                                                                                                                       (V(x_\tau,\tau)-V(y_\tau,\tau))]d\tau\}\rho_0(x_0,y_0)Dx_\tau Dy_\tau\rangle
$$
\vskip -1.2 truecm
$$
~~~~~~~~~~~~~~~~~~~~~~~~~~~~~~~~~~~~~~~~~~~~~~~~~~~~~~~~~~~~~~~~~~~~~~~~~~~~~~~~~~~t\!>\!\tau\!\geq\!0
$$
\vskip -.8 truecm
$$
\eqno{(9)}
$$
where $\rho_0$ is the initially given density matrix: $\rho_0(x,y)=\psi_0(x)\psi^{\!\!\!\ast}_0(y)$. On the rhs one can substitute
$$
\langle\exp\{-\frac{i}{\hbar}\int\displaylimits_0^t[V(x_\tau,\tau)-V(y_\tau,\tau)]d\tau\}\rangle
=\exp\{-\frac{1}{\hbar^2}\int\displaylimits_0^t g(x_\tau-y_\tau)d\tau\},
\eqno{(10)}
$$
which obviously follows from eq. (1) if we insert there $h(r,t)=-(1/\hbar)(\delta(r-x_\tau)-\delta(r-y_\tau))$ and use eq. (4). Thus we have path
integral representation for the density matrix at arbitrary time $t$:
\newpage
$$
\rho_t(x,y)=~~~~~~~~~~~~~~~~~~~~~~~~~~~~~~~~~~~~~~~~~~~~~~~~~~~~~~~~~~~~~
$$
$$
=\int\exp\{\int\displaylimits_0^t[i\frac{m}{2\hbar}({\dot x}^2_\tau-{\dot y}^2_\tau)-\frac{1}{\hbar^2}g(x_\tau-y_\tau)]d\tau\}
                                                                                                                         \rho_0(x_0,y_0)Dx_\tau Dy_\tau.
$$
\vskip -1.2 truecm
$$
~~~~~~~~~~~~~~~~~~~~~~~~~~~~~~~~~~~~~~~~~~~~~~~~~~~~~~~~~~~~~~~~~~~~~~~t\!>\!\tau\!\geq\!0
$$
\vskip -.9 truecm
$$
\eqno{(11)}
$$
 
 Differentiating both sides of eq. (11) by $t$, one gets the
 equation of motion for the density matrix in Gaussian white-noise
 potential (1):
 $$
 \frac{\partial}{\partial t}\rho_t(x,y)=\frac{i\hbar}{2m}(\frac{\partial^2}{\partial x^2}-\frac{\partial^2}{\partial y^2})\rho_t(x,y)
                                                                      -\frac{1}{\hbar^2}g(x-y)\rho_t(x,y).
 \eqno(12)
 $$
 
 This equation is sometimes called the master equation of the
 quantum noise theory. Our path integral method seems to be very
 effective for deriving the master equation even in more general
 external noises.
 
 The second term on the rhs of eq. (12) is a tipical damping
 term known from the quantum theory of reservoir effects (c.f.
 coarse grained approximation in ref. 3). This term cannot be reproduced
 by any given stochastic hamiltonian.
 
 Beside damping, a very peculiar feature of eq. (12) is that
 it produces \underline{mixed quantum state from a pure one} in a continuous
 manner. Exploiting the nature of this permanent quantum state
 mixing we shall construct the stochastic process for the evolution
 of the wavefunction itself.
 
 In order to make the calculation as simple as possible we
 suppose that
 $$
 g(r)=\frac{1}{2}A^2 r^2 +\mbox{higher order terms in }r;~~A=\mbox{const.}
 \eqno(13)
 $$
 and we shall neglect the ``higher terms'' by assuming that the width
 of the wavefunction will always be small enough.
 
 Thus, eq. (12) takes the form
 $$
 \frac{\partial}{\partial t}\rho_t(x,y)=\left[\frac{i\hbar}{2m}(\frac{\partial^2}{\partial x^2}-\frac{\partial^2}{\partial y^2})\rho_t(x,y)
                                                                      -\frac{A^2}{2\hbar^2}(x-y)^2\right]\rho_t(x,y).
 \eqno(14)
 $$
 If at time $t$ the particle is in a pure quantum state with the
 given wavefunction $\psi_t$ then [6]
 $$
 \rho_t(x,y)=\psi_t(x)\psi^{\!\!\!\ast}_t(y)
 \eqno(15)
 $$
 and eq. (14) yields
 $$
\rho_{t+\epsilon}(x,y)=~~~~~~~~~~~~~~~~~~~~~~~~~~~~~~~~~~~~~~~~~~~~~~~~~~~~~~~~~~~~~
$$
$$
=\left[1-\frac{\epsilon A^2}{2\hbar^2}(x-y)^2\right]
  \left[1+i\frac{\hbar\epsilon}{2m}\frac{\partial^2}{\partial x^2}\right]\psi_t(x)
  \left[1-i\frac{\hbar\epsilon}{2m}\frac{\partial^2}{\partial y^2}\right]\psi^{\!\!\!\ast}_t(y)                                                                           
\eqno{(16)}
 $$
 for infinitesimal $\epsilon>0$. Now, the rhs is not a single product like
 it was in eq. (15). Nevertheless, it can be decomposed into the
 sum of two such diadic terms:
 $$~~~~~~~~~~~~~~~~~~~~~~~~~~~~~~~~~~~~~~~~~~~~~~~~~~~~~~~~^\ast$$
 \vskip -1.0 truecm
 $$
\rho_{t+\epsilon}(x,y)=(1-\epsilon w)\psi_{t+\epsilon}(x)\psi^{\!\!\!\ast}_{t+\epsilon}(y) 
                                            +\epsilon w \psi_{t+\epsilon}(x) {\tilde\psi}_{t+\epsilon}(y) 
\eqno(17)
$$
where
 $$
w=\frac{A^2}{\hbar^2}\sigma^2_\psi
\eqno(18)
$$
is the mixing rate, the dominant wavefunction $\psi_{t+\epsilon}$ is
$$
\psi_{t+\epsilon}(x)=\{1-\frac{\epsilon A^2}{2\hbar^2}\left[(x-x_\psi)^2-\sigma^2_\psi\right]+i\frac{\epsilon\hbar}{2m}\frac{\partial^2}{\partial x^2}\}\psi_t(x),
 \eqno(19)
$$                               
and the contaminating wavefunction $\tilde\psi_{t+\epsilon}$ is
$$
\tilde\psi_{t+\epsilon}(x)=(\frac{x-x_\psi}{\sigma_\psi}+\frac{\epsilon A^2}{2\hbar^2}\frac{a_\psi^3}{\sigma_\psi})\psi_t(x).
 \eqno(20)
$$  
We introduced the following notations:
$$
x_\psi=\int x\vert\psi_t(x)\vert^2 dx,~~~~~~~~~~
$$
$$
\sigma^2_\psi=\int (x-x_\psi)^2\vert\psi_t(x)\vert^2 dx,
\eqno(21)
$$
$$
a^3_\psi=\int (x-x_\psi)^3\vert\psi_t(x)\vert^2 dx.
$$
It is easy to verify that $\psi_{t+\epsilon},~\tilde\psi_{t+\epsilon}(x)$ are normalized to the unity
and orthogonal to each other (in the lowest order of $\epsilon$, of course).

Now, let us read out the statistical meaning of eq. (17): in
an infinitesimally short time $\epsilon$, the quantum state $\psi_t$ of the particle
either evolves continuously into the neighbouring state $\psi_{t+\epsilon}$,
or, with the infinitesimal probability $\epsilon w$, jumps to the state
$\tilde\psi_{t+\epsilon}$, which is orthogonal to $\psi_{t+\epsilon}$.

Thus, the stochastic process governing the evolution of the
wavefunction is as follows. The wavefunction $\psi_t(x)$ of the given
particle satisfies the following non-linear equation of motion [9]:
$$
\frac{\partial}{\partial t}\psi_t(x)=i\frac{\hbar}{2m}\frac{\partial^2}{\partial x^2}\psi_t(x)
                                                                 -\frac{A^2}{2\hbar^2}\left[(x-x_\psi)^2-\sigma_\psi^2\right]\psi_t(x);
$$
\vskip -.2 truecm
$$
x_\psi=\int x\vert\psi_t(x)\vert^2 dx,~~~~~~~~~~~~~~~~~~~~~~~~~~~~~~~~~~~~~~~~~~~~
\eqno(22)
$$
\vskip -.2 truecm
$$
\sigma^2_\psi=\int (x-x_\psi)^2\vert\psi_t(x)\vert^2 dx,~~~~~~~~~~~~~~~~~~~~~~~~~~~~~~~~~~
$$
apart from discrete orthogonal jumps
\vskip -.2 truecm
$$
\psi_{t+0}=\frac{x-x_\psi}{\sigma_\psi}\psi_t(x)
\eqno(23)
$$
which occur at random in time. A jump is performed in an infinitesimal
interval $(t,t+\epsilon)$ with probability
\vskip -.2 truecm
$$
w\epsilon=\frac{A^2}{\hbar^2}\sigma^2_\psi\epsilon.
\eqno(24)
$$

By its construction, this stochastic process leads to the
same physical predictions in average as eq. (8) did. Namely, given
the initial density matrix $\rho_0(x,y)$, we can decompose it as
$$
\rho_0(x,y)=\sum_r p_r\psi_0^{(r)}\psi_0^{\!\!\!\ast\;(r)}(y)
\eqno(25)
$$
where $\psi_0^{(r)}$'s form an orthonormal system. Let us regard equality
(25) as if the particle were in the pure state $\psi_0^{(r)}$ with probability
$p_r$. Starting the stochastic process (22,23,24) from these
initial wavefunctions, each of them will give rise to the quantity
$$
\sum_r p_r\psi_t^{(r)}\psi_t^{\!\!\!\ast\;(r)}(y).
\eqno(26)
$$
The stochastic average of this expression over the histories $\psi_t^{(r)}$
is equal to $\rho_t(x,y)$.

We have to note that many other stochastic processes for $\psi_t$
can be constructed with the same $\rho_t(x,y)$. Nevertheless, we would
like to underline that the orthogonality of the stochastic jumps
(23) is very crucial from the viewpoint of measurement theory:
It can be shown that if we  know the wavefunction $\psi_t$ at $t=0$ then,
by means of von Neumann measurements [8], we can registrate all
stochastic jumps, \underline{without disturbing the measured particle}.  Thus,
in every moment, we are able to find out the wavefunction of the
system, if it satisfies indeed the eqs. (22,23,24). Of course, it
is not obvious at all how should we realize the proper measuring
apparatuses in practice.

Finally, we note the in the Langevin approach of quantum
damping [2] also stochastic process is constructed but this process
is related to the evolution of the density matrix, not to the
pure quantum state of the system.

I wish to thank to the authors of ref. 4 and also to Dr. P. 
Hrask\'o for illuminating discussions. 
 
\parskip 0truecm
\vskip 0.5 truecm
\noindent
REFERENCES
\vskip .2truecm

\noindent\hskip 10pt [1] W.H. Louisell and L.R. Walker, Phys. Rev. \underline{137}, B204 (1965)
\vskip .2 truecm

\noindent\hskip 10pt [2] M. Lax, Phys. Rev. \underline{145}, 110 (1966)
\vskip .2 truecm

\noindent\hskip 10pt [3] W.H. Louisell, Quantum Statistical Properties of Radiation\hfill\break 
$~~~~~~~~$(Wiley, New York, 1973)
\vskip .2 truecm

\noindent\hskip 10pt [4] F. K\'arolyh\'azi, A. Frenkel and B. Luk\'acs, in: Physics as\hfill\break  
$~~~~~~~~$Natural Philosophy, eds.  A. Shimony and H. Feshbach\hfill\break   
$~~~~~~~~$(MIT Press, Cambridge, 1982)
\vskip .2 truecm

\noindent\hskip 10pt [5] J.M. Ziman, Models of Disorder (Cambridge University Press,\hfill\break   
$~~~~~~~~$Cambridge, 1977)
\vskip .2 truecm

\noindent\hskip 10pt [6] L.D. Landau and E.M. Lifshits, Quantum Mechanics (Pergamon,\hfill\break   
$~~~~~~~~$Oxford, 1977) 
\vskip .2 truecm

\noindent\hskip 10pt [7] R.P. Feynman and A.R. Hibbs, Quantum Mechanics and Path\hfill\break   
$~~~~~~~~$Integrals (McGraw-Hill Book Company, New York, 1965)
\vskip .2 truecm

\noindent\hskip 10pt [8] J. von Neumann, Mathematische Grundlagen der Quantenmechanik\hfill\break   
$~~~~~~~~$(Springer Verlag, Berlin, 1932)
\vskip .2 truecm

\noindent\hskip 10pt [9] The equation (22) in itself, possesses solitonlike solutions.\hfill\break   
$~~~~~~~~$For similar mechanism see, e.g., L. Di\'osi, Phys. Lett. \hfill\break  
$~~~~~~~~$\underline{105A}, 199 (1984);. I. Bialynicki-Birula and J. Mycielski:\hfill\break  
$~~~~~~~~$Ann. Phys. \underline{100}, 62 (1976)
 \vskip .2 truecm
 
 \noindent\hskip 10pt [10] N. Gisin, Phys. Rev. \underline{A28}, 2891 (1983)                                                 

\end{document}